\documentclass[aps,physrev,twocolumn,superscriptaddress]{revtex4-2}

\usepackage{graphicx}% Include figure files
\usepackage{dcolumn}% Align table columns on decimal point
\usepackage{bm}% bold math
\usepackage{amsmath,amsfonts}
\usepackage[T1]{fontenc}
\usepackage{textcomp}
\usepackage{gensymb}
\usepackage{color}
\usepackage{xfrac}
\usepackage{soul}
\usepackage{amssymb}
\usepackage{hyperref}
\usepackage{url}
\usepackage{siunitx}

\hypersetup{
    colorlinks=false,
    linkcolor=blue,
    filecolor=magenta,      
    urlcolor=cyan,
}

\DeclareMathAlphabet{\mathsfit}{\encodingdefault}{\sfdefault}{m}{sl}
\SetMathAlphabet{\mathsfit}{bold}{\encodingdefault}{\sfdefault}{bx}{n}

\begin{document}

\title{Hardware Acceleration of Frustrated Lattice Systems using Convolutional Restricted Boltzmann Machine}

\author{Pratik Brahma}
\affiliation{Department of Electrical Engineering and Computer Sciences, University of California, Berkeley, California 94720, USA}

\author{Junghoon Han}
\affiliation{Department of Electrical Engineering and Computer Sciences, University of California, Berkeley, California 94720, USA}

\author{Tamzid Razzaque}
\affiliation{Department of Electrical Engineering and Computer Sciences, University of California, Berkeley, California 94720, USA}

\author{Saavan Patel}
\affiliation{Department of Electrical Engineering and Computer Sciences, University of California, Berkeley, California 94720, USA}
\affiliation{InfinityQ Technology Inc. Montreal, Quebec, Canada}

\author{Sayeef Salahuddin}
\email{sayeef@berkeley.edu}
\affiliation{Department of Electrical Engineering and Computer Sciences, University of California, Berkeley, California 94720, USA}
\affiliation{Materials Science Division, Lawrence Berkeley National Laboratory, California, 94720}

\date{\today}

\begin{abstract}
Geometric frustration gives rise to emergent quantum phenomena and exotic phases of matter.
While Monte Carlo methods are traditionally used to simulate such systems, their sampling
efficiency is limited by the complexity of interactions and ground-state properties. Restricted
Boltzmann Machines (RBMs), a class of probabilistic neural networks, offer improved
sampling by incorporating machine learning techniques. However, fully-connected
bipartite RBMs are inefficient for representing physical lattices with sparse interactions. To
address this, we implement Convolutional Restricted Boltzmann Machines (CRBMs) that
leverage translational symmetry inherent to lattices. Using the classical Shastry-Sutherland
(SS) Ising lattice, we demonstrate (i) CRBM formulation that captures SS interactions, and
(ii) digital hardware accelerator to enhance sampling performance. We simulate lattices with
up to 324 spins, recovering all known phases of the SS Ising model, including the long-range
ordered fractional plateau. Our hardware characterizes spin behavior at critical points and
within spin liquid phases. This implementation achieves 3–5 orders-of-magnitude speedup
(33 ns–120 ms) over GPU-based implementations. Moreover, the time-to-solution is within
two orders of magnitude of quantum annealers, while offering superior scalability, room-temperature
operation and reprogrammability. This work paves a pathway for scalable digital
hardware that embeds physical symmetries to enable large-scale simulations of material
systems.
\end{abstract}

\maketitle

\section{Introduction}
\label{sec:introduction}

\begin{figure*}
\includegraphics[width=0.8\linewidth]{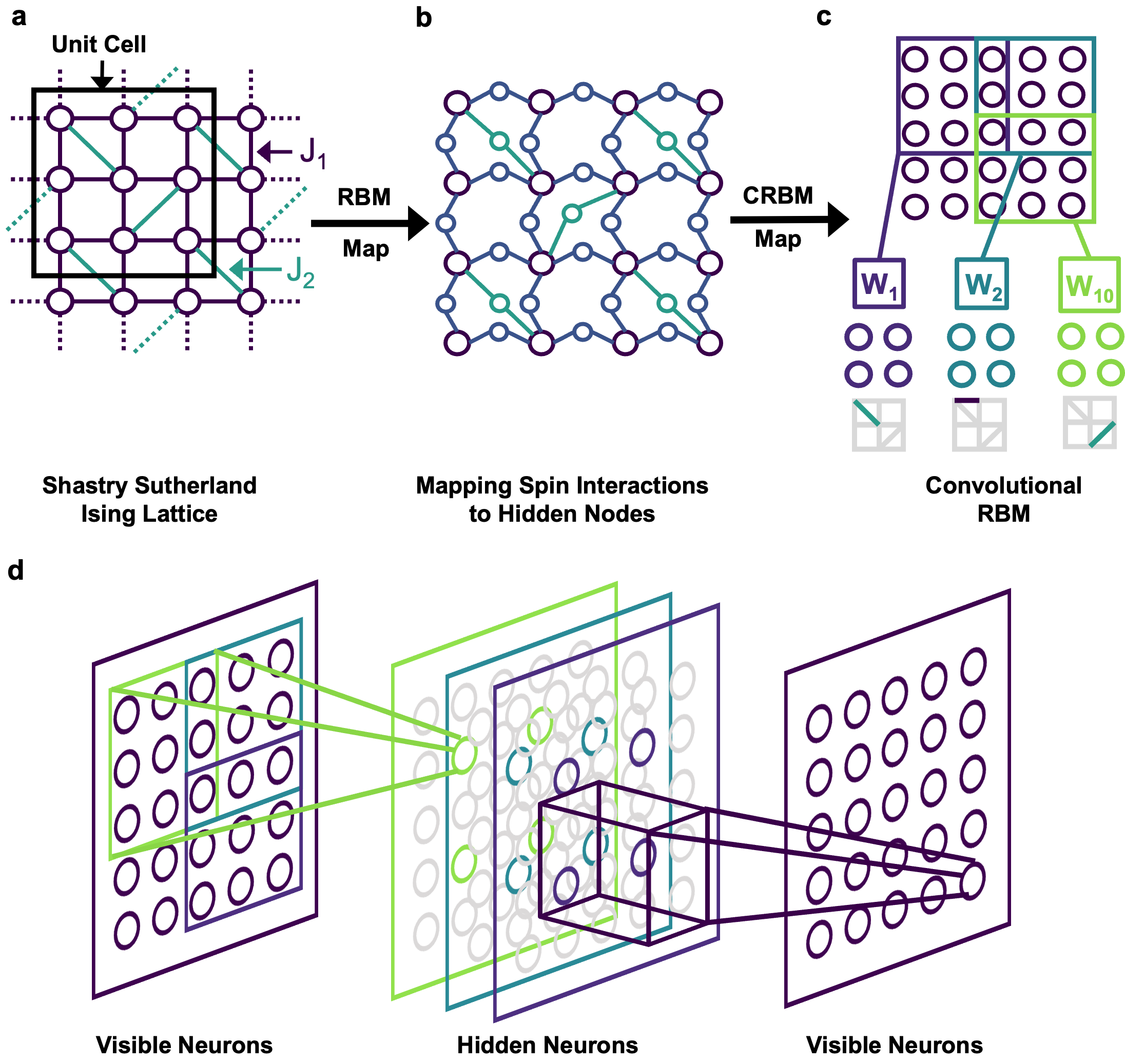}
\caption{\textbf{Convolutional Restricted Boltzmann Machine Mapping}: (a) The Shastry Sutherland (SS) Ising lattice structure with a $3\times3$ unit cell  (b) Mapping the spin interactions to hidden nodes using the formalism in \cite{Carleo2018}, resulting in a sparse RBM (c) Transforming the sparse RBM into a CRBM. Each colored convolutional filter represents a bond interaction within the 3X3 unit cell of the SS lattice. For this particular example, there are 10 bonds in the unit cell corresponding to 10 convolutional filters. (d) Gibbs sampling of the CRBM to obtain low lying energy spin configurations. The iterative cycle of updating visible and hidden neurons generates spin configurations that stochastically represent the Boltzmann distribution in Eq.\ref{Boltzmann_eq}.}
\label{Introduction Figure}
\end{figure*}

Frustration in spin-lattice \cite{Zhou2017,Carrasquilla2015,Arh2022,Kulbakov2021} gives rise to exotic emergent quantum phases in magnetic materials, as geometric constraints prevent the complete satisfaction of local interactions. 
Kagome and triangular lattices are prominent examples where geometric frustration results in exotic phases such as spin liquids.
 Solving such frustrated systems are inherently challenging due to the exponentially increasing state space and non-trivial ground state energy landscape. 
Traditionally, Monte carlo methods like simulated annealing \cite{LaCour2016}\cite{Hayami2022}, parallel tempering \cite{Huang2012}\cite{Hana2015} and cluster methods \cite{Barbu2014}\cite{Wolf89}, stochastically explore this exponential state space in search of ground states.
However, these conventional approaches face significant challenges, 
such as (i) disordered energy landscapes that lead to trapping in local minima, (ii) slow convergence in single-spin update methods due to correlated sampling, (iii) unreliable cluster updates in highly frustrated models due to their dependency on cluster size\cite{Cataudella94}, and (iv) critical slowing down near phase transitions\cite{Cossu_2018}.
In this regard, neural quantum states such as restricted boltzmann machines (RBMs)\cite{Fischer2012} have emerged as promising alternatives to alleviate some of these challenges.

RBMs are bipartite probabilistic neural networks that model the ground states of frustrated quantum systems through either training or direct mapping algorithms\cite{Nomura2017}\cite{Melko2019}.
For instance, RBMs have been shown to effectively capture the ground state of complex magnetic orders with large unit cells found in honeycomb lattices\cite{Li2021, Zou2022}.
In addition to quantum lattices, RBMs have been used as variational wavefuctions to minimize the quantum energy of material systems\cite{Kanno2021} and molecules\cite{nekrasov2020}. 
Moreover, many specialized digital implementations\cite{patel2020, patel2022, Kim2018} have been designed to further accelerate the Monte Carlo sampling procedure.
These hardware designs utilize RBMs' bipartite architecture, which enables simultaneous update of the entire spin lattice. 
However, for lattice systems that exhibit translational symmetry, the direct implementation of fully connected RBMs becomes inefficient. 
Specifically, the number of free parameters in an RBM scales as $L^4$, where $L$ is the system size, leading to inefficiencies in representing large systems. 
To address this, symmetries inherent in lattice systems can be exploited to design more efficient probabilistic neural networks.
By exploiting the translational symmetry, we derive the convolutional restricted Boltzmann machine (CRBM)\cite{Puente2020, Karthik2024}, where the convolutional filters correspond to the size of the unit cell and capture the physical interactions within that cell. 
This design offers two key advantages for lattice systems: (i) more efficient representations, since the number of parameters becomes independent of system size, and (ii) more efficient Monte Carlo sampling, leading to faster convergence and a larger number of uncorrelated samples\cite{Puente2020}.
As a result, CRBMs are more effective in modeling quantum systems with translational invariance. 
However, to the best of our knowledge, no hardware architectures have yet been developed based on CRBMs to provide hardware acceleration, complementing the algorithmic improvements offered by this architecture.

In this article, we demonstrate a digital implementation of CRBM on a Xilinx Field Programmable Gate Array (FPGA) to accelerate the solution of physical systems with translational symmetry.
Specifically, we show hardware acceleration by solving the classical Shastry Sutherland (SS) Ising lattice, as illustrated in Fig.\ref{Introduction Figure}a)\cite{Dublenych2012}:
\begin{eqnarray}
\mathcal{H} \left( \mathbf{\sigma}\right) = J_1\sum_{\langle i,j \rangle} \sigma_i\sigma_j + J_2\sum_{\langle \langle i,j \rangle \rangle}\sigma_i\sigma_j + h_z\sum_i\sigma_i
\label{SS eqn}
\end{eqnarray}
This Hamiltonian is particularly suited to showcase the efficiency of our hardware for several reasons. 
First, the Shastry-Sutherland lattice is a geometrically frustrated system, where competition between the nearest-neighbor ($J_1$) and next-nearest-neighbor ($J_2$) interactions leads to a variety of exotic phases.
These include a long-range ordered 1/3 magnetization plateau when $J_1 \sim J_2 > 0$\cite{Dublenych2012} and an emergent classical spin liquid phase induced by lateral confinement\cite{Brahma2024}.
Second, this system has previously been solved using quantum annealers, providing a benchmark for evaluating our digital implementation's performance\cite{Kairys2020}. 
Third, this frustrated lattice model and its variants explain the magnetism in materials such as rare-earth tetraborides $\text{RB}_4$, where $\text{R} = {\text{Tm}, \text{Er}, \text{Ho}}$\cite{Yoshii2008, Kim2009, Trinh2018, Siemensmeyer2008}, thus highlighting potential material realizations.
Finally, this system can be directly mapped onto CRBM filters (see Fig.\ref{Introduction Figure}a-c), bypassing the need for variational training and emphasizing the hardware acceleration of our architecture.
Motivated by these reasons, we employ our CRBM hardware to stochastically explore the energy landscape and identify ground state spin configurations for varying $J_1$ and $J_2$ interactions and in the presence and absence of lateral confinement.
We simulate lattice systems with 324 logical spins and successfully recover all known phases of the SS Ising model in the thermodynamic limit.
In addition to phase recovery, we analyze the spin structure factor of ground state configurations at critical transitions and within the classical spin liquid phase, thus establishing connections to diffuse neutron scattering experiments.
Our results demonstrate  3-5 orders of acceleration (in 33 ns to 120 ms) compared to GPU-based algorithms and performance within 2 orders of magnitude (in 2 ms) relative to quantum annealers.
Furthermore, we extract scaling power laws that describe the time-to-convergence behavior across different phases.
Our work paves the way for designing digital hardware that efficiently captures the inherent symmetries of physical systems, particularly CRBMs designed for systems with translational symmetry.
Given that RBMs are suitable for solving quantum lattices and real materials, the algorithmic improvements offered by CRBMs, and the demonstrated hardware acceleration, our approach enables large-scale simulations of complex material systems.

\section{Methods}
\label{sec:methods}

This work implements the Restricted Boltzmann Machine (RBM)\cite{Hinton2012}, a probabilistic graph neural network, to model the Shastry-Sutherland (SS) Ising lattice. 
The RBM represents a probability distribution over a bipartite graph comprising two sets of nodes, (i) visible neurons ($\mathbf{v}$), which represent physical variables of the system such as lattice spins\cite{Li2021, Zou2022} or molecular orbitals\cite{nekrasov2020}, and (ii) hidden neurons ($\mathbf{h}$) that capture the interactions between these physical variables.
The probability distribution of the RBM is then mathematically described as:
\begin{eqnarray}
\text{E}\left(\mathbf{v}, \mathbf{h} \right) = \sum_i a_iv_i + \sum_j b_jh_j + \sum_{ij}v_iW_{ij}h_j
\end{eqnarray}
\begin{eqnarray}
\mathbb{P}\left(\mathbf{v}, \mathbf{h} \right) = \frac{1}{Z}\exp(-\beta \text{E}\left(\mathbf{v}, \mathbf{h} \right))
\label{Boltzmann_eq}
\end{eqnarray}
where $a_i$ and $b_j$ are biases for the $i^{\text{th}}$ visible neuron and $j^{\text{th}}$ hidden neuron respectively, and $W_{ij}$ represents the weight interaction between the $i^{th}$ visible and $j^{th}$ hidden neuron. 
The parameter $\beta = 1/T$ denotes the inverse-temperature, a hyperparameter that controls the escape rate from local minima during Monte Carlo sampling.
Finally,  the constant $Z$ ensures that $\mathbb{P}(\mathbf{v}, \mathbf{h})$ is normalized.
In quantum lattice and molecular systems, the RBM represents the variational ground state function.
Solving such systems requires training to optimize the weights and biases of the RBM that minimize the quantum energy.
In contrast, for classical ising lattices-specifically the SS Ising lattice-the RBM represents the entire energy landscape, with Monte Carlo sampling providing a stochastic representation of this landscape.
Additionally, we derive the Convolutional Restricted Boltzmann Machine by incorporating the translational symmetry present in many crystalline solids and lattice structures into the RBM.
This architecture leverages weight sharing, making the number of learned parameters independent of the system size, in contrast to RBMs\cite{Puente2020}.
The CRBM architecture is illustrated in Fig.\ref{Introduction Figure}c.
The following sections detail: (i) an analytical mapping from the classical SS lattice model to a CRBM, (ii) the sampling procedure for generating configurations that represent the energy landscape, and (iii) the hardware architecture of our CRBM implementation.

\subsection{Mapping to CRBM}
\label{subsec:mapping}

\begin{figure*}
\includegraphics[width=0.9\linewidth]{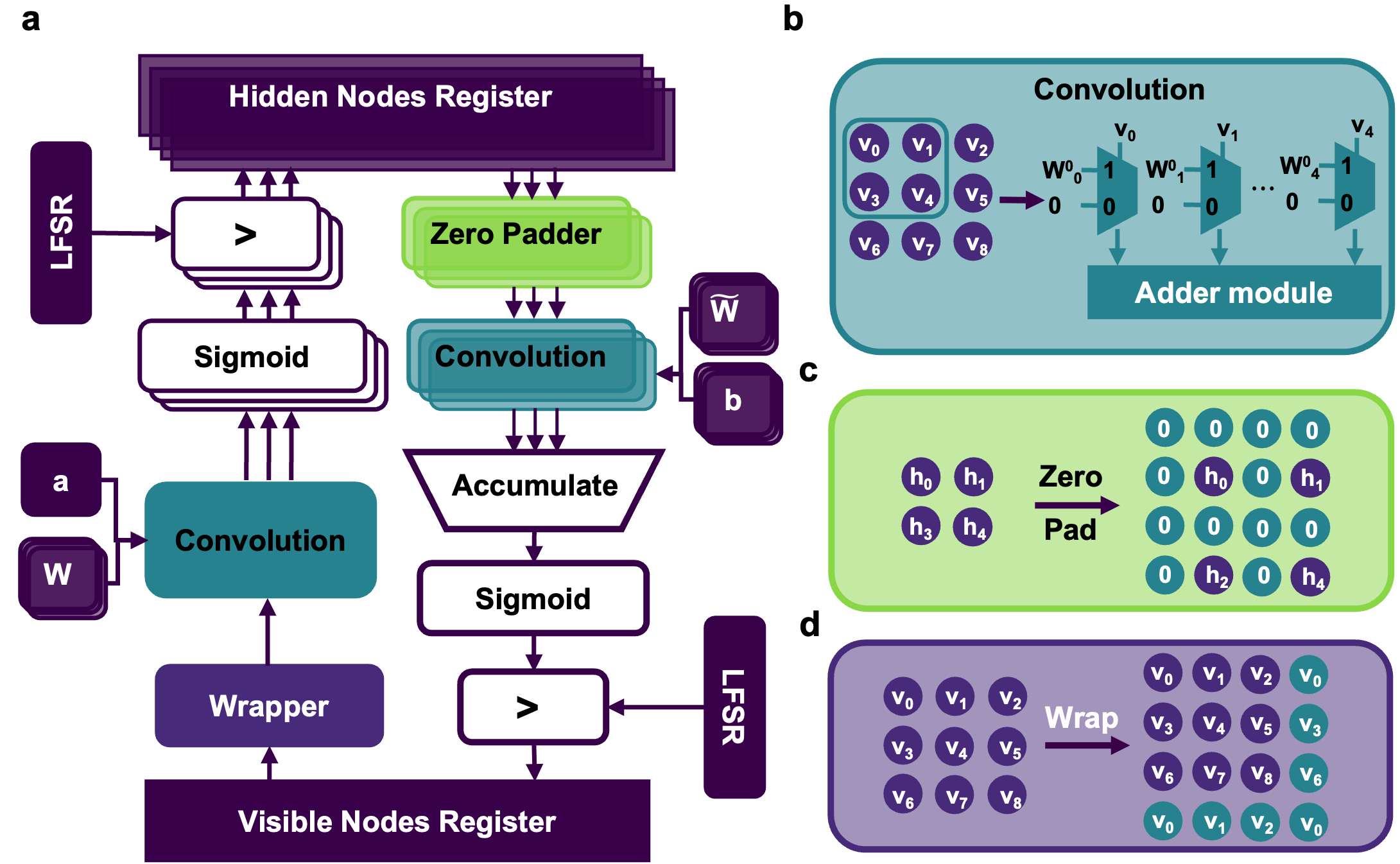}
\caption{\textbf{Hardware Implementation of CRBM}: (a) The spatially parallelized, two-stage pipelined, and logically synthesized CRBM hardware accelerator. The two columns of operations between the node registers represent the two parallel processing pipelines. (b) Convolution operation that models the spin interactions. This is implemented using a mask-and-accumulate operation with multiplexer-based bitwise multiplication, enabling resource-efficient and high-speed sampling. (c) Zero padding ensures that the reconstructed visible neuron retains the original dimensions. (d) The wrapping module emulates periodic or open boundary conditions of the lattice.}
\label{Hardware Architecture}
\end{figure*}

Here, we present the direct mapping of the translationally symmetric SS ising lattice to a CRBM.
Each spin of the lattice is mapped to a visible neuron and each interacting bond is mapped to a hidden neuron as shown in Fig.\ref{Introduction Figure}b. 
To derive the analytical mapping equations, we equate the Boltzmann distribution of the physical lattice model to the marginal distribution of the corresponding sparse RBM over visible neurons\cite{Carleo2018}:
\begin{eqnarray}
\frac{1}{{Z}_{phys}}\exp(-\beta \mathcal{H}\left(\mathbf{v} \right)) = \sum_\mathbf{h} \mathbb{P}_{RBM} \left(\mathbf{v}, \mathbf{h}\right)\nonumber \\
= \frac{1}{Z}\exp\left(-\beta\sum_i a_i v_i \right) \prod_j \cosh(\beta b_j + \beta\sum_i v_i W_{ij} )
\end{eqnarray}
This equation must hold for all $2^n$ possible combinations of the visible neurons, where $n$ is the number of visible neurons. 
One solution includes mediating each spin interaction through a hidden neuron\cite{Carleo2018}, described by:
\begin{eqnarray}
\exp(J v_i v_j) = C\sum_{h = \pm 1} \exp(\hat{J_i} v_ih + \hat{J_j} v_j h) \nonumber\\= 2C\cosh(\hat{J_i}v_i + \hat{J_j} v_j)
\end{eqnarray}
\begin{eqnarray}
\hat{J_i} = \frac{1}{2}\text{arccosh}\left(\exp(2|J |)\right) \\
\hat{J_j} = \frac{1}{2}\text{sgn}(J)\text{arccosh}\left(\exp(2|J |)\right)\\
C = \frac{1}{2}\exp(-|J|)
\end{eqnarray}
This approach maps all spin interactions to hidden neurons, converting the lattice structure to a sparse RBM. 
Given that the SS lattice exhibits discrete translation symmetry $\mathcal{H}(\mathcal{T}_2\mathbf{\sigma}) = \mathcal{H}(\sigma)$, where $\mathcal{T}_2$ is a translation operator which shifts the spin state by 2 lattice sites in either direction, the weights of the resulting sparse RBM are also translationally invariant as depicted in Fig.\ref{Introduction Figure}b. 
Consequently, the RBM weights are unique up to the unit cell of the SS Lattice (size $3 \times 3$) where the unit cell is shown in Fig.\ref{Introduction Figure}a. 
Incorporating the $\mathcal{T}_2$ symmetry and the analytical mapping equations, we can reformulate the previous sparse RBM as a two-dimensional CRBM\cite{Puente2020, Karthik2024}:
\begin{eqnarray}
    \text{E}\left(\mathbf{V}, \mathbf{H}\right) = a\sum_{ij} \mathbf{V}_{ij} + \sum_{k}\text{Tr}\left(\left(\mathbf{H}^k\right)^T \left(\mathbf{V} * \tilde{W}^k \right) \right)
\end{eqnarray}
\begin{eqnarray}
    \tilde{W}^k = \begin{cases}
      \frac{1}{2} \text{arccosh}\left(\beta J_1 \right)\mathcal{M}_k & k \in \left[0,7\right] \\
      \frac{1}{2} \text{arccosh}\left(\beta J_2 \right)\mathcal{M}_k & k \in \left\{8,9\right\}
      \end{cases}
\end{eqnarray}
In this formulation, each bond within the SS lattice unit cell is labeled from 0 to 9.
Here, $\mathbf{V}$ is a two-dimensional matrix of visible neurons, and $\mathbf{H^k}$ is a two-dimensional matrix of hidden neurons in the $k^{th}$ layer, $*$ denotes cross-correlation operator and $\tilde{W}^k$ represents the weight matrix for the $k^{th}$ convolutional layer. 
The masking matrix $\mathcal{M}_k$ selects visible neurons connected along bond $k$ within the unit cell.
Thus, the CRBM is uniquely defined by the visible bias $a$ and the convolutional weight filters $\tilde{W}^k$, obtained from the $J_1$ and $J_2$ parameters of the SS ising model.

\subsection{Sampling of CRBM}
\label{subsec:sampling}

Gibbs sampling\cite{Geman1984} allows the CRBM to stochastically explore the energy landscape of the mapped lattice.
Since the CRBM represents the Boltzmann distribution over this landscape, low-lying energy states correspond to high-probability regions.
Consequently, the Gibbs sampling algorithm spends more time in the low-lying energy states, facilitating convergence to the classical Hamiltonian ground states.
Gibbs sampling operates on the conditional distributions of visible and hidden neurons as follows:
\begin{eqnarray}
    \mathbb{P}_{RBM}\left( \mathbf{H}^k_{ij}= 1 | \mathbf{V} \right) = \text{sigmoid}\left( \mathbf{V} * \tilde{W}^k\right)_{ij}
\label{vis_to_hid_eq}
\end{eqnarray}
\begin{eqnarray}
    \mathbb{P}_{RBM}\left( \mathbf{V}_{ij} = 1| \mathbf{H} \right) = \text{sigmoid}\left(\sum_k \mathbf{\tilde{H}}^k * \tilde{W}^{Rk}\right)_{ij}
\label{hid_to_vis_eq}
\end{eqnarray}
This Gibbs sampling process for the CRBM involves two key steps.
First, all hidden neurons are simultaneously sampled based on the visible neurons using Eq.\ref{vis_to_hid_eq} and the convolutional filters $W^k$.
Specifically, for the SS lattice, the convolution filters are $3 \times 3$ in size, and the cross-correlation operation involves a stride of 2.
Next, all visible neurons are simultaneously sampled based on the updated hidden neuron using the conditional probability given in Eq.\ref{hid_to_vis_eq}.
In this step, two new variables are introduced: $\mathbf{\tilde{H}}^k$, the zero-padded version of the $k^{\text{th}}$ layer hidden neuron matrix $\mathbf{H}^k$, and $\tilde{W}^{Rk}$, the reversed version of the original weight matrix $\tilde{W}^k$, where rows and columns are flipped.
Zero padding (as shown in Fig.\ref{Hardware Architecture}c) compensates for the size reduction caused by strided convolution when updating hidden from visible neurons, ensuring the reconstructed visible neuron matrix matches the original dimensions. 
The reversal of convolutional filters arises from the specific indexing order of visible and hidden neurons.
These visible and hidden neuron updates are iteratively applied (as shown in Fig.\ref{Introduction Figure}d), where each consecutive sample instance stochastically represents the underlying probability distribution.

\subsection{CRBM Digital Hardware}
\label{subsec:hardware}

The CRBM graph architecture contains several salient features that make it an ideal candidate for hardware acceleration.
These features include (i) conditional independence of visible and hidden neurons, enabling the simultaneous update of the entire lattice system, (ii) the integration of memory (weights and biases) and the computation (sigmoidal activation) of the CRBM within the same digital fabric, exemplifying in-memory computing \cite{Karunaratne2020, Mannocci2023, Sebastian2020} and (iii) bitwise convolution which facilitates efficient hardware design by replacing the traditionally expensive matrix multiplication operations in specially designed neural network hardware\cite{Li2023, Elbtity2020}.
These characteristics collectively enhance the computational efficiency and scalability of our CRBM-based hardware implementation.

Fig.\ref{Hardware Architecture} illustrates the spatially parallelized, two-stage pipelined, and logically synthesized CRBM hardware accelerator. 
This accelerator comprises two pipeline stages corresponding to forward sampling (visible to hidden neuron update) and backward sampling (hidden to visible neuron update). 
In the forward sampling stage, samples are produced according to the conditional probability defined in Eq. \ref{vis_to_hid_eq}.
This process starts with a wrapper module that simulates periodic or open boundary conditions of the lattice (see Fig.\ref{Hardware Architecture}d.)
Next, a convolution operation is applied to the wrapped visible nodes using the user-specified weight filters and bias.
This operation employs a mask-and-accumulate operation with multiplexer-based bitwise multiplication, as illustrated in Fig.\ref{Hardware Architecture}b.
The resulting convolved output is then passed through a sigmoid function, which is approximated using a logical lookup table to minimize hardware resource consumption.
Finally, the hidden neurons are updated by comparing the sigmoid output with a random number generated by a 16-bit linear feedback shift register (LFSR), exemplifying probabilistic sampling of binary random variables.
Similarly, the backward sampling stage generates samples based on the conditional probability specified in Eq.\ref{hid_to_vis_eq}.
This stage begins with a zero padder that ensures the reconstructed visible neuron retains the original dimensions (see Fig. \ref{Hardware Architecture}c)
The same convolution operation is then applied across the $N$ independent hidden neuron layers, where $N$ represents the total number of weight filters.
The outputs from these $N$ layered convolution operations are then summed over via an accumulator and a sigmoid function is subsequently applied.
Finally, the sigmoid output is compared with a random number generated from LFSR to update the visible neuron register.

All these mathematical operations are executed in parallel across the entire digital logic area, with computations performed in place at each visible and hidden node. 
The pipeline stages, separated by registers for the visible and hidden nodes, facilitate instruction-level parallelism by enabling the simultaneous execution of two independent sampling cycles.
One cycle is denoted as an even cycle when sampling begins from a randomly initialized visible node register, whereas the other cycle is denoted as an odd cycle when the sampling begins from the hidden node register.
The hardware implementation comprises up to 1134 nodes (324 visible + 810 hidden) and operates at a clock frequency of 30MHz, consuming approximately 5W.
However, due to the dual independent sampling cycles enabled by pipelining, samples are produced effectively at 60 MHz.
A more detailed description of the hardware acceleration is given in the Supplementary Information.

In this work, the CRBM hardware is utilized as an inference accelerator, with the host machine transferring problem parameters to the FPGA. 
These parameters include weight filters, biases, and lattice boundary conditions, as well as sampling-related settings such as the number of sampling cycles. 
Communication between the FPGA and the host machine, including writing problem parameters and retrieving generated sample configurations, is conducted via an x16 PCIe interface\cite{Preuber2014}.
Beyond its role as an inference engine, the hardware can also serve as a variational wavefunction to accelerate variational Monte Carlo\cite{Puente2020, Karthik2024} algorithms.
In this context, the host machine iteratively refines the variational wavefunction by updating the CRBM's weight filters based on the previously generated samples that statistically represent the variational wavefunction.

\section{Results}
\label{sec:results}

\begin{figure*}
\includegraphics[width=0.9\linewidth]{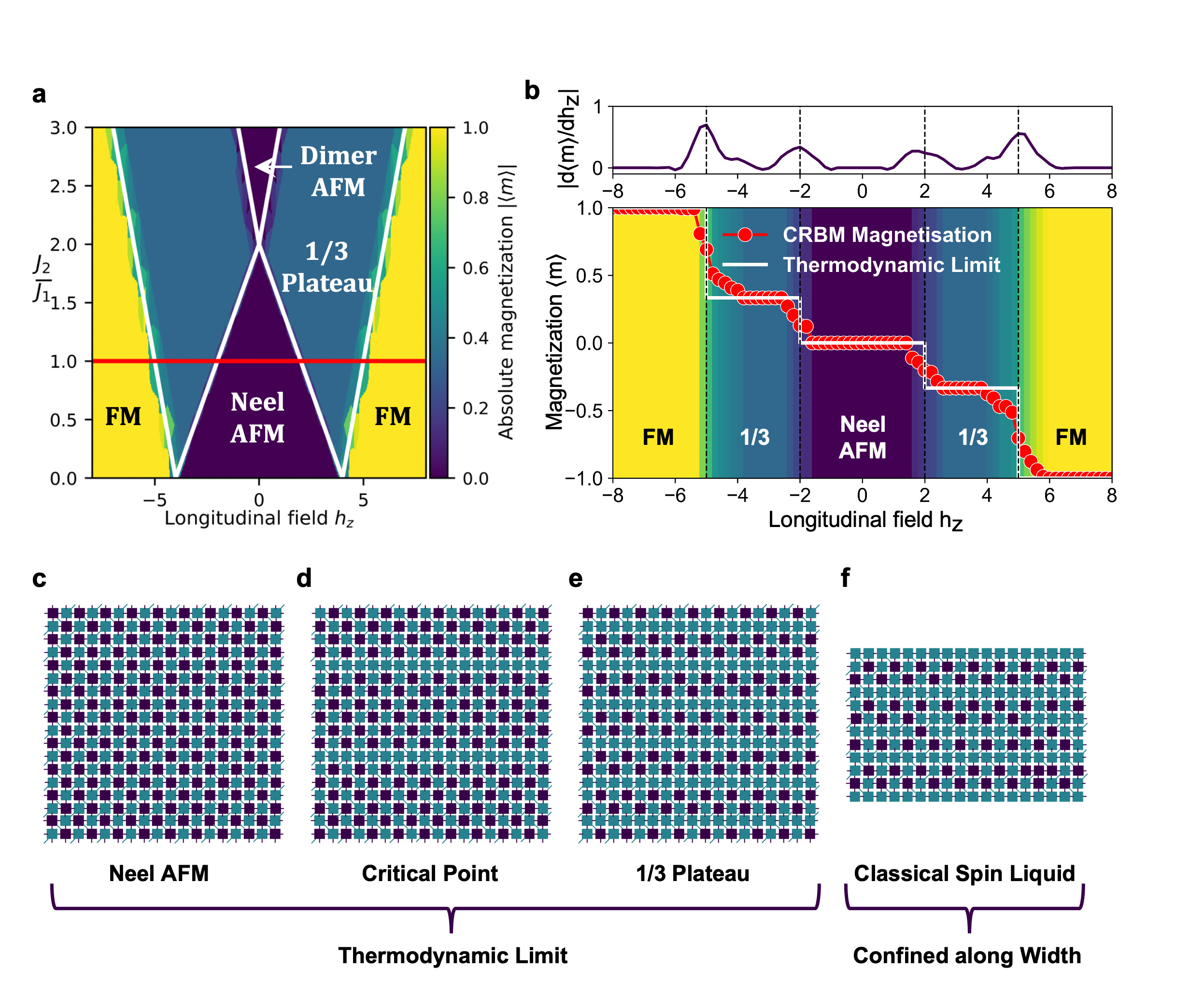}
\caption{\textbf{Phase Diagram and ground state spin configurations}:  (a) Magnetization Phase diagram of the SS lattice of size $18\times18$ obtained using the CRBM hardware accelerator. The thermodynamic limit was approximated by imposing periodic boundary conditions. (b) Average magnetization curve as a function of longitudinal field along $J_2/J_1 = 1.0$. The broadening near transition points arises from finite-temperature simulations and the limited precision of fixed-point convolutional filter representations. Additionally, the peaks in the first derivative of the curve coincide with the theoretical critical points. (c) Neel AFM spin configuration (d) Spin configuration at the critical point  $J_2/J_1 = 1.0, h_z = 2.0$ and (e) Spin configuration corresponding to the 1/3-magnetization plateau. (f) Classical spin liquid state emerging from lateral confinement\cite{Brahma2024} in an SS lattice of size $12\times18$. The problem parameters were $J_2/J_1 = 1.5, h_z = 4.5$. The blue squares represent -1 and the green squares represent 1.}
\label{Phase Diagram}
\end{figure*}

In this work, we map the SS Ising lattice onto our CRBM hardware implementation and obtain ground state spin configurations from the generated samples.
Each single logical spin is mapped to a single visible node, and interactions among the logical spins are represented by convolutional filters.
This mapping results in 324 visible nodes and 810 hidden nodes for an SS Ising lattice of size $18 \times 18$.
The following sections present various results related to ground-state spin configurations, static structure factors, and hardware acceleration achieved by our CRBM digital implementation.
In Section "Phase Diagram", we recover all magnetization plateaus in the thermodynamic limit and compare our hardware-based results to exact analytical solutions \cite{Dublenych2012}\cite{Dublenych2014}.
Additionally, we observe the emergence of classical spin liquid configurations due to lateral confinement\cite{Brahma2024}.
In Section "Static Structure Factors", we compute and analyze static structure factors for the obtained spin configurations, including those at critical points and for laterally confined lattices.
Finally, in Section "Hardware Acceleration", we demonstrate acceleration in time-to-solution (TTS) by our CRBM digital implementation, comparing it to an equivalent GPU-based algorithmic approach.

\subsection{Phase Diagram}
\label{subsec:thermodynamic_limit}

The SS Ising lattice of lattice size $18 \times 18$ is mapped onto the CRBM digital implementation with periodic boundary conditions to simulate the thermodynamic limit\cite{Fisher1970, Tavares2015}.
Optimizing the parameter $\beta$ in the Boltzmann distribution (see Eq.\ref{Boltzmann_eq}) allows us to confidently explore the extensive problem space defined by $J_1$, $J_2$, $h_z$, despite the limited precision inherent in the fixed-point representation used for convolutional filters and the pseudo-random number generator.
We obtain an ensemble of low-lying energy states for  $J_1 = 1.0$, $J_2 \in [0,3]$ and $h_z \in [-8, 8]$.
For each problem configuration, the CRBM is executed 10 times, with each run generating approximately $3 \times 10^6$ samples from which the lowest energy spin configurations are extracted.

Fig.\ref{Phase Diagram}a illustrates the phase diagram generated from ground state energy solutions obtained using the CRBM, which agrees with theoretical predictions\cite{Dublenych2012}.
All long-range ordered magnetization plateaus are recovered, and phase transitions are identified by the white lines in Fig.\ref{Phase Diagram}(a).
These phase transition lines are determined by differentiating the magnetization phase diagram with respect to $h_z$ $\left( \text{i.e } \partial m/ \partial h\right)$ and selecting regions where the derivative peaks\cite{}. 
This procedure is further illustrated in Fig.\ref{Phase Diagram}b, which plots the magnetization along  $J1 = 1.0$ and shows that the derivative peaks correspond to the theoretical critical phase transition points.
The smooth transitions in magnetization observed in our results, in contrast to the abrupt transitions predicted theoretically, can be attributed to thermal broadening introduced by Boltzmann distribution sampling in the CRBM and the limited accuracy introduced by fixed point precision of convolutional filters, sigmoid lookup tables, and accumulator.

Finally, the CRBM Monte Carlo samples enable a real-space analysis of the ground state spin configurations, as shown in Fig.\ref{Phase Diagram}c-e for various longitudinal field ($h_z$) values.
Fig.\ref{Phase Diagram}c displays a typical antiferromagnetic (Neel AFM) spin configuration at $J2 = 1.0, hz = 0.0$, while Fig.\ref{Phase Diagram}e illustrates the long-ranged ordered 1/3 fractional state at $J2 = 1.0, hz = 3.0$.
This state is characterized by a minimal repeating unit consisting of a UUD (Up-Up-Down spin) configuration, producing two interleaving AFM rows sandwiched by ferromagnetic rows.
Fig.\ref{Phase Diagram}d shows the spin configuration at the critical phase transition ($J_2 = 1.0$, $h_z = 2.0$) between Neel AFM and the 1/3 fractional plateau.
This configuration appears as a mixture of the AFM and fractional state, with blocks of interleaving AFM rows sandwiched by ferromagnetic rows.
Sampling at this critical point generates an ensemble of ground state solutions, of which only one representative state is presented.
Additionally, our hardware implementation supports open boundary conditions along the lateral direction.
This feature enables the study of classical spin liquid phases arising from lateral confinement in the Shastry–Sutherland Ising lattice\cite{Brahma2024}.
Fig.\ref{Phase Diagram}f presents a classical spin liquid state for a lattice of width 12 at $J2 = 1.5, hz = 4.5$, which theoretically corresponds to a correlated dimer classical spin liquid state belonging to the $W = 3n$ group\cite{Brahma2024}.
Similar to the critical point configuration, sampling this phase yields an ensemble of ground state solutions, further confirming the nature of classical spin liquids\cite{Gingras2011, Ramirez1999, Myers2000}. 
These states can be further characterized by examining their static structure factors, as discussed in the next section.

\begin{figure*}
\includegraphics[width=0.85\linewidth]{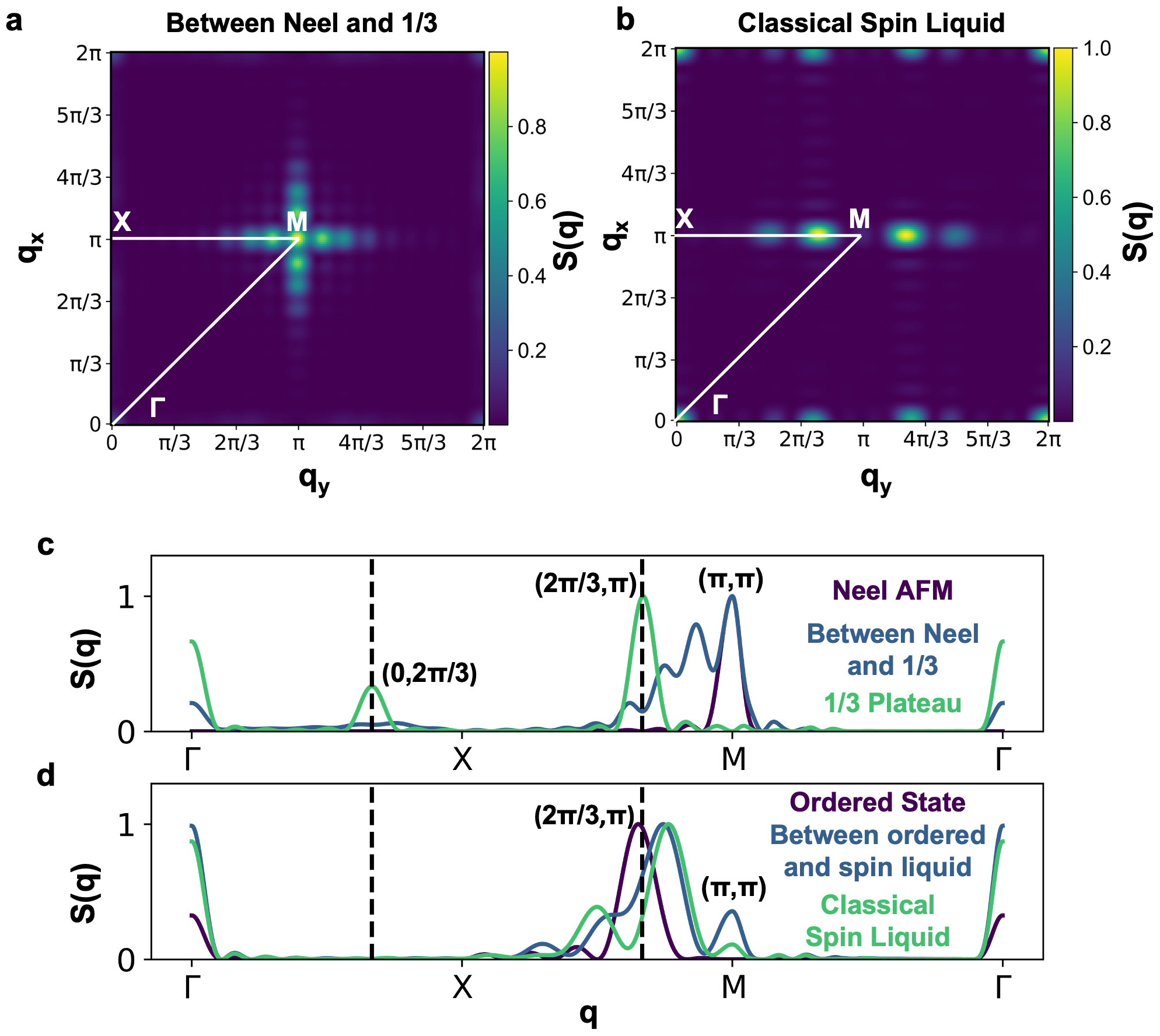}
\caption{\textbf{Static Structure factor of various spin configurations}: (a) Static structure factor of the spin configuration at the critical transition between Neel AFM and 1/3 fractional plateau. The structure factor appears as a broadened superposition of peaks characteristic of both the AFM and fractional plateau phases. (b) Structure factor of the classical spin liquid phase at $J_2/J_1 = 1.5, h_z = 4.5$. (c) A cut along the high symmetry points of (a) illustrates the transition of structure factor from Neel AFM to 1/3 fractional plateau. At the critical point, the peaks broaden over a range of momenta depicting the destruction of long-range order-a signature of critical phase transitions. (d) A cut along the high symmetry points of (b), showing the transition from an ordered state at $J_2/J_1 = 1.5, h_z = 2.5$ to a classical spin liquid phase, with the transition occuring at $h_z = 3.5$. At the transition point and within the classical spin liquid phase, $S(\vec{q})$ exhibits broadening over a wide range of momenta, reflecting macroscopic degeneracy—a defining characteristic of classical spin liquids \cite{Ramirez1994}.}
\label{Static Structure Factor}
\end{figure*}

\subsection{Static Structure Factors}
\label{subsec:static_structure_factors}

Experimental techniques such as diffuse neutron scattering play a crucial role in studying magnetic ordering in complex systems, including spin glasses\cite{Mydosh2015}, low-dimensional magnets\cite{Coldea2001}, and quantum spin liquids\cite{Balents2010}.
These experiments probe the static structure factor $S(\vec{q})$ via the Fourier decomposition of the spin correlations.
Following the approach in \cite{Kairys2020}, we establish a connection between our theoretical analysis and experimental techniques by computing the static structure factor.
$S(\vec{q})$ is defined in terms of the two-point spin correlation function:
\begin{eqnarray}
S(\vec{q}) = \sum_{i,j}\langle \sigma_i \sigma_j \rangle e^{i\vec{q} \cdot (\vec{R}_i - \vec{R}_j)}
\end{eqnarray}
where $\vec{R}_{i,j}$ is the relative position between spins at sites $i$ and $j$, and $\vec{q}$ is the momentum vector.
To compute $S(\vec{q})$ at a specific problem parameter, we follow the same simulation protocol as in the phase diagram analysis, generating an ensemble of low-energy spin configurations. 
This order parameter captures the distribution of magnetic order and spin correlations in the reciprocal space, further characterizing the previously described real-space spin configurations.

Fig.\ref{Static Structure Factor}a,c illustrate the static structure factor for spin configurations at the critical point ($J_2 = 1.0, hz = 2.0$) between the Neel AFM phase and the 1/3 fractional plateau.
The structure factor at this transition appears as a broadened superposition of the peaks characteristic of both the AFM and the fractional plateau phases.
As shown in Fig.\ref{Static Structure Factor}c, the AFM phase exhibits a sharp peak at $(\pi, \pi)$ corresponding to the alternating antiferromagnetic rows, while the 1/3 plateau features a distinct peak at $(2\pi/3, \pi)$, corresponding to its characteristic UUD spin configuration\cite{Dublenych2012}.
In contrast, at the critical point, the peaks broaden over a range of momenta between the peak positions characteristic of the two ordered phases (AFM and 1/3 plateau). 
This broadening depicts the destruction of long-range order, a signature of critical phase transitions\cite{Sethna2006}.

Fig.\ref{Static Structure Factor}b,d depicts the static structure factor for spin configurations corresponding to the classical spin liquid phase ($J_2 = 1.5, hz = 4.5$) in laterally confined lattices of width 12.
Specifically, Fig.\ref{Static Structure Factor}d shows the transition of the structure factor from an ordered phase to a  classical spin liquid.
For lattices with widths following the $3n$ series, we observe an ordered UUD spin configuration at $J2 = 1.5, hz < 3.5$ and a  classical spin liquid at $J2 = 1.5, hz > 3.5$, with the critical point occurring at $h_z = 3.5$\cite{Brahma2024}.
In the ordered UUD phase, the static structure factor exhibits sharp peaks at $(2\pi/3, \pi)$.
In contrast, increasing the longitudinal field to and beyond the critical point ($h_z = 3.5$) broadens these peaks over a wide range of momenta, indicating macroscopic degeneracy and spin fluctuations characteristic of classical spin liquids\cite{Ramirez1994}.
Notably, the classical spin liquid phase shows a magnetic preferential direction along a slightly shifted $(2\pi/3, \pi)$, which arises due to frustration and lateral confinement effects.
Moreover, both at the critical point and in the classical spin liquid phase, peaks are observed at the $\Gamma$ point, likely due to the presence of ferromagnetic rows as evidenced by the real-space analysis in Fig\ref{Phase Diagram}f.
These results demonstrate that our hardware is capable of exploring and analyzing critical phenomena in frustrated lattice systems and exotic phases in spin liquids, while also providing significant hardware acceleration. 
This acceleration is detailed in the following section.

\begin{figure*}
\includegraphics[width=0.9\linewidth]{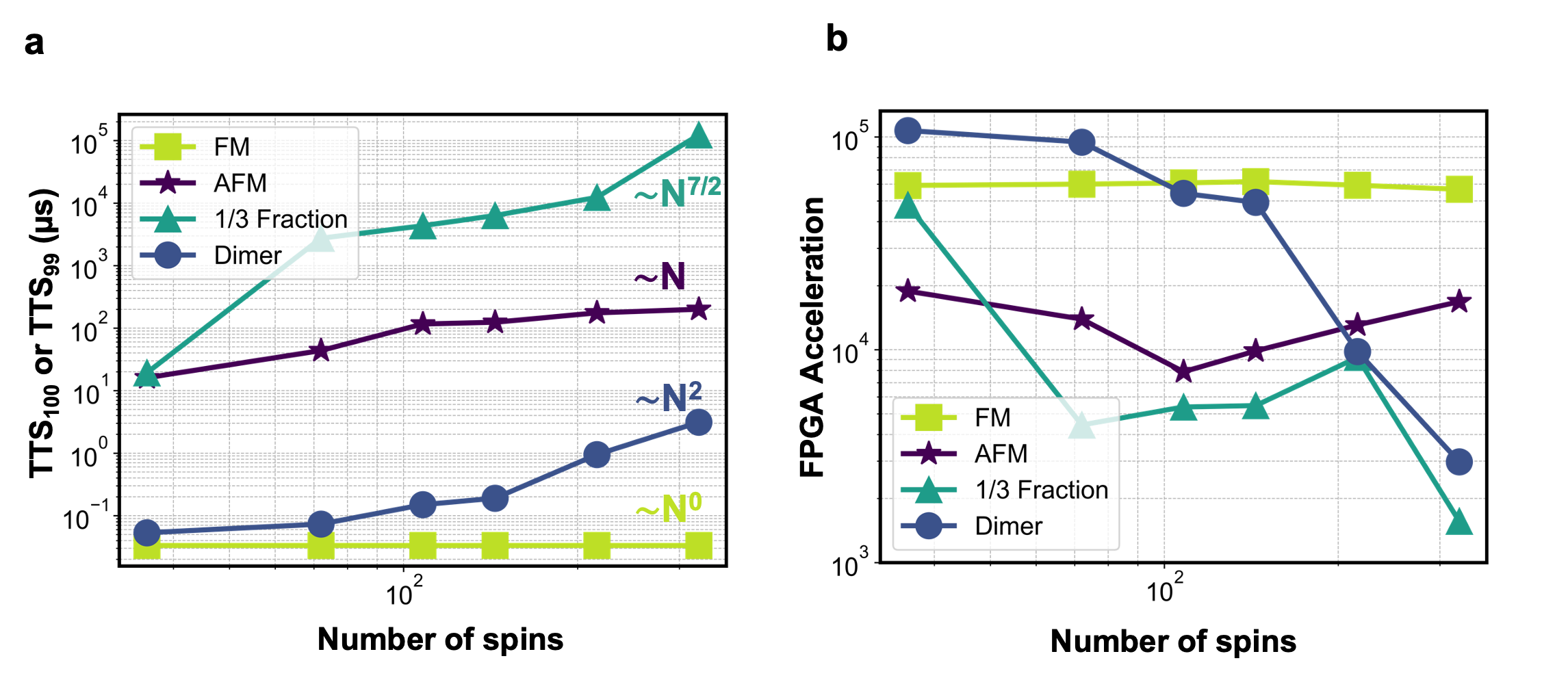}
\caption{\textbf{Hardware Acceleration across various thermodynamic magnetic phases}: (a) TTS Scaling of the hardware across different thermodynamic magnetization phases. The FM ($J_2/J_1 = 1.0, h_z = 7.0$) and the dimer phase ($J_2/J_1 = 2.5, h_z = 0.0$) are the easiest to sample due to the presence of large longitudinal bias and macroscopic degeneracy, respectively. In contrast, Neel AFM ($J_2/J_1 = 1.0, h_z = 0.0$) and 1/3 fractional ($J_2/J_1 = 0.5, h_z = 4.0$) are much more challenging to sample due to the finite degeneracy of ground state solutions and the presence of long-range magnetic order. (b) Hardware acceleration of our digital implementation compared to GPU-based CRBM sampling algorithm.}
\label{Scaling TTS}
\end{figure*}

\subsection{Hardware Acceleration}
\label{subsec:scaling}

We use time-to-solution (TTS)-a common metric for benchmarking stochastic algorithms\cite{King2018, Cai2020, Hamerly2019}-to demonstrate hardware acceleration on the SS Ising lattice problem.
TTS is defined by:
\begin{eqnarray}
    \text{TTS}_l = \text{T} \frac{\log{\left(1 - l/100\right)}}{\log{\left(1 - p_{gnd}\right)}}
\end{eqnarray}
where $l$ depicts the desired confidence level (in percent), $T$ is the average time required for the stochastic algorithm to reach a ground state solution, and $p_{gnd}$ is the probability of finding the ground state in a single run.
In this study, we focus on $\text{TTS}_{99}$ ($l = 99$), which represents the time needed to achieve a solution with $99\%$ confidence level. 
We estimate $p_{gnd}$ by running the hardware ten times independently and calculating the fraction of total runs that successfully reach the ground state solution.
When the hardware achieves a 100\% success rate ($p_{gnd} = 1$), we denote the time-to-solution as  $\text{TTS}_{100}$
Using this metric, we investigate the TTS scaling for lattice sizes ranging from $6\times6$ (36 spins) to $18 \times 18$ (324 spins) across various thermodynamic magnetization phases, including the ferromagnetic (FM), Néel antiferromagnetic (AFM), 1/3 fractional plateau, and dimer phases.
Finally, we analyze the scaling behavior of the TTS curves to determine the relative difficulty of sampling each thermodynamic phase with our CRBM-based approach.
 
Fig.\ref{Scaling TTS}a illustrates the TTS scaling for various thermodyamic magnetic phases.
Notably, this scaling depends on three factors: (i) the efficiency of our CRBM-based Monte Carlo sampling, (ii)  the inherent inaccuracies arising from the fixed-point representation of the convolutional filters and the hardware's random number generator, and (iii) the magnetic order and degeneracy of the ground state spin configurations.
We observe that the ferromagnetic (FM) phase is the easiest to sample, requiring only 33 ns, and its scaling remains independent of system size ($N^0$).
This behavior can be attributed to the strong longitudinal bias, which drives the entire lattice into the FM configuration simultaneously via the CRBM sampling. 
The dimer phase, which is sampled in around 5 $\mu$s, is the next easiest, owing to its large manifold of degenerate solutions. 
However,  its TTS scaling ($\sim N^2$) is worse than the more challenging Neel AFM phase ($\sim N$), a consequence of the hardware's fixed point representation of interactions that sacrifices precision for speed.
Exotic phases, such as spin configurations at the critical point and the classical spin liquid phase in laterally confined lattices, exhibit similar scaling behavior due to their macroscopic degeneracy and the breakdown of long-range order\cite{Ramirez1994, Sethna2006}. 
Finally, AFM (in $200$ $\mu$s) and the 1/3 fractional phase (in $120$ ms) are the most difficult to sample, requiring the longest TTS because of their long-range order and finite ground-state.

Figure.\ref{Scaling TTS}(b) illustrates the hardware acceleration achieved across various thermodynamic magnetic phases relative to a GPU-based CRBM sampling approach.
Here, acceleration is defined as the ratio of the TTS measured on our CRBM hardware to the TTS on the GPU implementation.
We observe that the ferromagnetic (FM) and Néel antiferromagnetic (AFM) phases exhibit consistent accelerations of approximately five and four orders of magnitude, respectively, across different lattice sizes.
In contrast, the 1/3 fractional and dimer phases show a decrease in acceleration with increasing lattice size, but still provide over three orders of magnitude in speedup at the largest system sizes.
This hardware acceleration can be attributed to several key features of our CRBM hardware design: (i) expensive multiply-and-accumulate (MAC) operations are replaced by bitwise multiplexers owing to the bit representation of visible and hidden neurons; (ii) a fixed-point representation of convolutional filters is utilized, resembling quantization of weights in deep neural networks for fast and efficient inference\cite{Wei2024}; and (iii) translational symmetry is leveraged to enable efficient sampling of large systems.
Our hardware operates within two orders of magnitude of quantum annealer's performance (2 ms)\cite{Kairys2020}  while offering several distinct advantages: (i) scalability and room-temperature operation enabled by mature digital technology, (ii) reprogrammability that allows convolutional filters to map arbitrary interactions within the unit cell—unlike quantum annealers, which require the lattice to be embedded in specific qubit connectivity graphs\cite{Zbinden2020}, and (iii) seamless integration with a CPU host via a PCIe interface.
This capability facilitates accelerated solutions for quantum material systems possessing translational symmetry, where the CRBM can serve as a variational wavefunction\cite{Puente2020, Karthik2024}.
Consequently, these results demonstrate a promising pathway for designing digital hardware that captures essential symmetries in quantum systems, enabling large-scale simulations of quantum materials.

\section{Conclusion}
\label{sec:conclusion}

To conclude, we have demonstrated a digital hardware implementation of a convolutional restricted Boltzmann machine (CRBM) that leverages the translational symmetry inherent in diverse quantum systems, including crystals, frustrated lattices, and magnetic materials.
Specifically, we focus on the classical Shastry–Sutherland Ising lattice to showcase the hardware's capabilities in analyzing spin configurations at critical points, exploring emergent classical spin liquid phases via lateral confinement, and deriving phase diagrams in the thermodynamic limit.
By calculating structure factors, we provide a means of comparing our theoretical results with experimental observations from techniques such as neutron diffuse scattering.
Additionally, our hardware architecture programmed on FPGA demonstrates a three to five-order-of-magnitude speedup (33 ns to 120 ms) over GPU-based sampling methods across various thermodynamic magnetic phases.
This acceleration stems from key design features, including bitwise multiply-and-accumulate (MAC) operations, fixed-point weight representations, and the exploitation of translational symmetry.
Notably, our performance is within two orders of magnitude of quantum annealers such as D-Wave (2 ms)\cite{Kairys2020} while offering significant advantages in scalability and reprogrammability.
Moreover, the seamless integration of our hardware with a host CPU allows the acceleration of variational Monte Carlo algorithms for solving quantum materials, where the CRBM hardware can serve as a variational wavefunction\cite{Puente2020, Karthik2024}.
By combining the enhanced efficiency of CRBM sampling with hardware architectures
that exploit symmetry, this work opens a promising route for large-scale simulations of
quantum materials and frustrated lattices in a faster and more efficient manner.

\begin{acknowledgments}
This work was supported by ASCENT, one of six centers in JUMP, a Semiconductor Research Corporation (SRC) program sponsored by DARPA.
\end{acknowledgments}

% Appendix with Supplementary Material
\appendix

\section{Detailed Description of CRBM Hardware Accelerator}
\label{app:hardware}

In this section, we present the technical details of our CRBM hardware accelerator.
The primary objective of this accelerator is to significantly reduce the run-time required to reach the ground-state solution of CRBM.
The hardware design is implemented at the Register Transfer Level (RTL) using Verilog as hardware description language (HDL) and mapped onto a Field-Programmable Gate Array (FPGA). 
Specifically, we used the Virtex UltraScale+ FPGA (VCU118) from Xilinx–AMD.
This FPGA features state-of-the-art 14nm/16nm FinFET process technology, dynamic power management, and integrated Gen3 x16 PCIe blocks, making it a highly efficient platform for our implementation.
Our experimental setup includes this FPGA in conjunction with a server equipped with an 11th Gen Intel Core i9-11900K processor (3.50 GHz) and 135 GB of RAM. 
We employ Xilinx's Vivado toolchain for FPGA synthesis and programming.

\begin{figure}
\includegraphics[width=0.9\columnwidth]{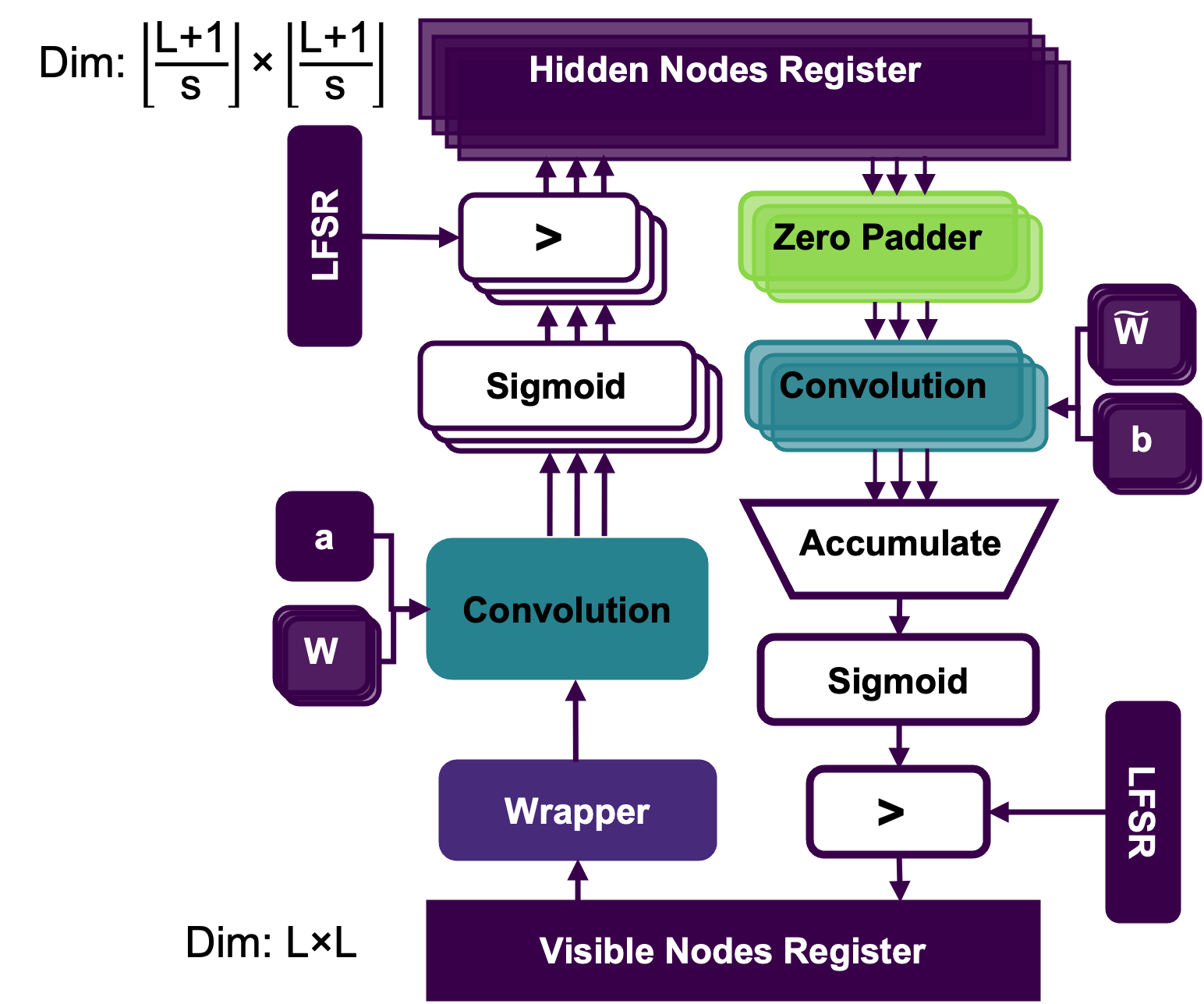}
\caption{\textbf{Hardware Architecture of CRBM}: Pipelined hardware architecture as depicted in the main text. Here, $s$ represents the stride of the convolution.}
\label{SuppFig1}
\end{figure}

The hardware architecture, as illustrated in Fig.\ref{SuppFig1}, maps each step of the CRBM to the respective hardware modules. 
The hardware is pipelined with two stages: forward and reverse. The forward stage follows the sampling logic from visible nodes to hidden nodes (stages 1 to 4). 
The reverse stage contains the sampling logic from hidden nodes to visible nodes (stages 5 to 9).

\subsection{Visible Node Registers}

The 2D visible node layer is stored in a single register, with each node represented by a single bit due to its binary nature.
This approach minimizes the use of Look-Up Table (LUT) resources on the FPGA.
The size of the lattice is indicated as $L\times L$, indicating $L$ rows and $L$ columns of visible nodes, resulting in a total of $L^2$ visible nodes. Each node represents a single binary value.
Consequently, the visible node register consists of $L^2$ bits.

\subsection{Wrapper Module}

Wrapping is implemented to incorporate periodicity into the convolution logic.
Assume that the filter size is $M \times M$. 
When periodicity is enabled in the column direction, the first $M-1$ columns are copied to the last column indices.
Conversely, when periodicity is disabled, $M-1$ columns of zeros are inserted. 
The same procedure applies to the row direction. The output dimension of the wrapper is $(L + M - 1) \times (L + M - 1)$.
The wrapping module takes the visible node register and periodicity signal as inputs, modifying the corresponding columns and rows by either copying data or zero-padding according to the boundary condition.

\subsection{Convoluter - Forward}

The forward convolution module implements the convolution operation required to sample hidden nodes from visible nodes.
The convolution process consists of an element-wise matrix multiplication between the filter and the visible nodes at the current position, followed by an accumulation operation (MAC).
This operation is performed iteratively with a stride, shifting the filter to the next position in both the row and column directions.
This mapping pattern continues until the indices exceed the lattice boundaries.
The procedure remains identical for all convolutional filters, with the number of output groups equal to the number of distinct convolutional filters.
This implementation exploits spatial parallelism by integrating multiply-and-accumulate logic directly at each node position.
Identical logic units are instantiated at positions spaced by the stride distance in all directions. 
Consequently, all convolution computations are executed within a single, spatially parallelized combinational logic module, thereby allowing this computation to be done in a single clock cycle.
The filter and bias values are provided by the user software via the PCIe communication interface.

\subsection{Sigmoid and LFSR-Forward}

The sigmoid modules are synthesized with configurable input and output bit widths, which determine the hardware's precision.
The input of each module is the result of the convolution and the output is the corresponding sigmoid value.
Internally, each sigmoid module contains a precomputed Look-Up Table (LUT) that functions as a key-value mapping of input-output pairs.
The module selects the closest precomputed sigmoid value based on the specified precision parameters.

The Linear Feedback Shift Register (LFSR) module is synthesized using a set seed value.
The module's internal register, initialized with the seed value, is shuffled every cycle to produce randomized bits. 
The LFSR output is converted to a value between 0 and 1.
For $N$ convolutional filters, the design includes $N$ sigmoid modules and $N$ LFSR modules. Both the sigmoid function and the LFSR use 16-bit representations for floating-point values ranging from 0 to 1.

The sigmoid output is compared against the corresponding LFSR-generated value: if the sigmoid value exceeds the LFSR output, the hidden node is assigned a value of 1; otherwise, it is assigned 0.
This logic completes the first pipeline stage.

\subsection{Hidden Node Registers}

For $N$ filters, the forwarded convolution results in $N$ hidden node groups. 
Each group has dimensions $\lfloor (L + 1)/s \rfloor \times \lfloor(L + 1)/s \rfloor$. 
Consequently, the hidden node register stores a total of
$N \times \lfloor(L + 1)/s\rfloor \times \lfloor(L + 1)/s \rfloor$ bits.

\subsection{Zero Padder}

A zero-padding technique is applied to ensure that the reverse sampling process (hidden-to-visible) maintains the same dimensions as the original visible node layer.
This is achieved by inserting zeros between hidden nodes in all directions.
Similarly to the wrapping step, zero-padding also involves copying the last columns and rows to the beginning of the corresponding dimensions.
If periodicity is enabled, we copy the hidden node values
along with padded zeros. 
If periodicity is disabled, only zeros are added to the first row and column.
The hardware maintains an array of zeros with designated slots for hidden node values. 
These values are inserted in a spatially parallel manner.
For $N$ convolutional filters, the design includes $N$ zero-padding modules, each corresponding to a group of hidden nodes.

\subsection{Convoluter - Reverse}

The convoluter module used for reverse sampling is identical to the one employed in the forward direction.
In this stage, the module receives flipped filter weights as inputs, which the user software provides via PCIe.
Therefore, this design allows the hardware to flexibly support different filter weights for forward and reverse convolution operations.
For $N$ filters, the hardware implements $N$ reverse convoluter modules.

\subsection{Accumulator}

The output of the previous $N$ convoluter modules is accumulated in this module.
Since accumulation occurs at corresponding positions among the 
$N$ filter groups, a simple combinational logic efficiently sums these $N$ values.
Following accumulation, the $N$ filter groups are aggregated into a single output group. 
Additionally, a visible bias is applied in this module, with bias values provided by the user.
The design includes an option to apply odd and even biases, allowing different bias values to be assigned to odd and even columns.

\subsection{Sigmoid and LFSR Reverse}

The sigmoid and LFSR modules used in the reverse direction are identical to those employed in the forward direction.
The outputs of these modules determine the next visible node values.
This stage marks the completion of the second pipeline stage, thereby concluding a full cycle of sampling.

\subsection{Input and Output (I/O) Programming Logic}

The host machine and FPGA communicate via an x16 PCIe interface.
We implement the Input and Output (I/O) logic of the PCIe through an open source module named Xillybus.
The Input/Output (I/O) logic for PCIe is implemented using Xillybus, an open-source module that provides both an FPGA IP core and a driver for the host PC's operating system.
Xillybus offers customized bundles for different FPGA models, facilitating seamless integration.
Although our hardware need not communicate large data within each time step, the host machine and FPGA run on different clock frequencies, producing a clock domain crossing.
To address this, we employ a First-In-First-Out (FIFO) module, enabling efficient sequential communication between the host and FPGA.
The Xillybus driver creates two device files for FPGA communication: one for writing and one for reading. The user software interacts with these files as follows: (i) Write (CPU to FPGA): weights, flipped weights, biases, lattice sizes (visible layer dimensions), periodicity settings, and a signal to clear the last hidden row (for applications requiring clamping to zero) (ii) Read (from FPGA to CPU): visible node values of each cycle.
After retrieving visible node values from the FPGA, the user software computes the energy of the visible nodes according to the system being studied.

\subsection{Pipelining}

Since the hardware is pipelined into two stages (forward and reverse), each complete sampling of a visible node set requires two cycles. 
Simultaneously, the design allows two independent samplings to occur in parallel.
At each cycle, the visible node register and hidden node registers store values corresponding to the two different sampling chains.
These registers alternate between two sampling chains, enabling independent samplings on odd and even cycles.
This dual-chain execution effectively doubles the expected throughput, enhancing overall performance.

\subsection{Dimension and Precision}

The maximum lattice dimension that fits on our FPGA is $18 \times 18$, corresponding to 324 visible nodes and 810 hidden nodes.
Our hardware implementation uses a $3\times3$ filter size, 10 filters, and 16-bit floating-point precision for the sigmoid and LFSR modules.
For convolution calculations, we employ 10-bit precision, allocated as 1 sign bit, 5 integer bits, and 4 floating-point bits. The Accumulator module employs additional 2 bits to mitigate overflow.

\subsection{Clock Frequency and Critical Path}

The hardware was synthesized on the FPGA with a clock frequency of 30 MHz. 
For each sampling chain, the visible node updates its values every two cycles, completing a full sampling sequence in approximately 66.66 nanoseconds.
Since there are two independent chains, the hardware produces one unique sample every 33.33 nanoseconds.
The critical path, which dictates the clock frequency, is in the forward pipeline stage (visible to hidden).
A significant portion of the FPGA area is allocated to the sigmoid and LFSR modules within this stage.
Reducing the precision bits of these modules can alleviate the critical path constraint to some extent. 
However, this trade-off comes at the cost of reduced accuracy or an increased number of sampling iterations required to reach the ground-state solution.
Our current choice of precision bits represents an optimized balance between computational accuracy and timing constraints.

% Supplementary Figure for Appendix

% Bibliography
\bibliography{references}

\end{document}